\documentclass[prl, twocolumn, nobalancelastpage, nobibnotes, superscriptaddress]{revtex4}

\begin{document}
\title{Note on Non-relativistic proof of the spin-statistics connection in the Galilean frame}

\author{E. C. G. Sudarshan}
\affiliation{Center for Particle Physics, Department of Physics, University of Texas, Austin, TX 78712}

\author{Anil Shaji}
\affiliation{Center for Statistical Mechanics, Department of Physics, University of Texas, Austin, TX 78712}
\email{shaji@physics.utexas.edu}

\begin{abstract}
We reply to the critique by Pucchini and Vucetich of our construction of a non-relativistic proof of the spin-statistics connection using $SU(2)$ invariance and a Weiss-Schwinger action principle.  
\end{abstract}


\maketitle

In a recent paper \cite{puccini04}, Puccini and Vucetich have argued that a non-relativistic proof of the spin-statistics connection cannot be obtained in the Galilean frame. The paper is a critique of a proof suggested by us in \cite{shaji03}. The main assertion in \cite{puccini04} is that Hermitian field operators are incompatible with Galilean invariance if the fields are not massless. hermiticity and Galilean invariance; both are indeed required of the field operators in our proof presented in \cite{shaji03}. 

The argument of Puccini and Vucetich is centered on the transformation properties of the field operator $\xi_{\lambda}({\bf x}, t)$ under the Galilei group:
\begin{equation}
  \label{eq:trans1}
  U_g \xi_{\lambda}({\bf x}, t)U^{-1}_g = e^{\frac{i}{\hbar} m \gamma(g; {\bf x}, t)} \sum_{\lambda'} D^{s}_{\lambda \lambda'} (R^{-1}) \xi_{\lambda'} ({\bf x'}, t)
\end{equation}
where $\xi_{\lambda}$ is a field operator with spin $s$ and $\lambda = -s, \ldots , s$, $D^{s}_{\lambda \lambda'}$ is the $(2s+1)$ dimensional unitary representation of the rotation group and $\gamma(g; {\bf x}, t)=\frac{1}{2} {\bf v}^2 t + {\bf v} \cdot {\bf x}$. The transformation properties for the field $\xi_{\lambda}^{\dagger}$ follow from equation (\ref{eq:trans1}):
\begin{equation}
  \label{eq:trans2}
  U_g \xi_{\lambda}^{\dagger}({\bf x}, t)U^{-1}_g = e^{-\frac{i}{\hbar} m \gamma(g; {\bf x}, t)} \sum_{\lambda'} D^{s}_{\lambda' \lambda} (R) \xi_{\lambda'}^{\dagger} ({\bf x'}, t).
\end{equation}

The inequivalent Bargmann phase \cite{bargmann54} picked up by $\xi_{\lambda}$ and $\xi_{\lambda'}$ under the transformation is used by Puccini and Vucetich to conclude that ``{\em no Galilean field operator of non-zero mass can he hermitian}''. 

In the usual (complex) realization of the extended Galilei group the finite transformations are a mixture of real and complex transformations. The generators of rotations and space translations $J_{jk}$ and $P_j$ are pure imaginary and the corresponding finite transformations $e^{i {\bf J} \cdot {\bf \theta}}$ and $e^{i{\bf P}\cdot{\bf a}}$ are real. On the other hand the generators of time translation and boosts, $H$ and $G_j$ are usually chosen to be real so that the corresponding finite translations are imaginary. In the extended Galilei group $M$ is taken to be a non-negative real number and so it also generates transformations that are not real. If we want the field operators to be hermitian then we want all the finite transformations to be real too so that under the action of an element of the Galilei group the real components of the field gets mapped on to real components. This can be accomplished by doubling the number of components of $\xi_{\lambda}$ and choosing $M$, $H$ and $G_j$ as follows:
    \begin{eqnarray}
      \label{eq:realgen}
      M & = & m \left( \begin{array}{cc} 0 & - i \\ i & 0 \end{array} \right), \nonumber \\ \nonumber \\
      H & = &\frac{p^2}{2M}=  \frac{p^2}{2m} \left( \begin{array}{cc} 0 & - i \\ i & 0 \end{array} \right), \nonumber \\ \nonumber \\
      G_j & = & M q_j = m q_j \left( \begin{array}{cc} 0 & - i \\ i & 0 \end{array} \right).
    \end{eqnarray}
Doubling the number of components for $\xi_{\lambda}$ also means that the rotation and space translation generators which are already pure imaginary have to be expanded appropriately. This choice representation allows us to keep the field operators hermitian while at the same time keeping the Lagrangian density constructed out of $\xi_{\lambda}$ Galilean invariant. Using this choice of representation of the extended Galilei group we can again go through with the proof constructed in \cite{shaji03} and apply it to massive Galilean fields also. The choice of Galilei group generators that has to be made to keep the fields hermitian may be cumbersome for most computations. However it shows that all the assumptions necessary for our construction of the proof can in principle be justified for massive fields that transform according to representations of the (extended) Galilei group.

The essential point is that a field (of any spin) can be made to carry an additional charge by doubling the components while still keeping them real. The 'mass' $M$, which may be considered as just another charge, can also be accommodated in an identical fashion. To lament over the Bargmann phase due to $M$ and not worry about any other charge in relation to the proof of the spin-statistics connection stems from assigning $M$ a special status over any other charge that may be relevant to the fields that are being considered.

\end{document}